\documentclass[pra,aps,amssymb,twocolumn,noshowpacs]{revtex4-1}
\usepackage{color}
\usepackage{graphicx}
\usepackage{hyperref}
\usepackage{amsmath}
\usepackage{latexsym}
\usepackage{amssymb}
\usepackage{mathrsfs}
\usepackage{mathtools}
\usepackage{layout}
\usepackage{verbatim}
\usepackage{bm}
\usepackage{amsfonts,epsfig}

\begin{document}
\title{Universal Barenco quantum gates via a tunable non-collinear interaction}

\date{\today}
\author{Xiao-Feng Shi}
\affiliation{School of Physics and Optoelectronic Engineering, Xidian University, Xi'an 710071, China}

\begin{abstract}
The Barenco gate~($\mathbb{B}$) is a type of two-qubit quantum gate based on which alone universal quantum computation can be achieved. Each $\mathbb{B}$ is characterized by three angles ($\alpha,\theta$, and $\phi$) though it works in a two-qubit Hilbert space. Here we design $\mathbb{B}$ via a non-collinear interaction $V|r_1r_2\rangle\langle r_1r_3|+$H.c., where $|r_i\rangle$ is a state that can be excited from a qubit state and $V$ is adjustable. We present two protocols of $\mathbb{B}$. The first~(second) protocol consists of two~(six) pulses and one~(two) wait period(s), where the former causes rotations between the qubit states and excited states, and the latter induces gate transformation via the non-collinear interaction. In the first protocol, the variable $\phi$ can be tuned by varying phases of external controls, and the other two variables $\alpha$ and $\theta$, tunable via adjusting the wait duration, have a linear dependence upon each other. Meanwhile, the first protocol can give rise to the CNOT and Controlled-Y gates. In the second protocol, $\alpha,\theta$, and $\phi$ can be varied by changing the interaction amplitudes and wait durations, and the latter two are dependent on $\alpha$ non-linearly. Both protocols can also lead to another universal gate when $\{\alpha,\phi\}=\{1/4,1/2\}\pi$ with appropriate parameters. Implementation of these universal gates is analyzed based on the van der Waals interaction of neutral Rydberg atoms. 
\end{abstract}

\maketitle

\section{introduction}

The data processing in computers involves many gate operations, which is also the case for quantum computation although the latter works in fundamentally different ways. The information processing in quantum computing can be understood as a series of unitary operation upon a given input state~\cite{Nielsen2000,Williams2011}. If a set of quantum gates can represent an arbitrary unitary operation, it is a universal set~\cite{Deutsch1989}. Study of universal sets of quantum gates has been a focus for decades~\cite{Barenco1995,Sleator1995,PhysRevA.51.1015,Loss1998,PhysRevLett.89.097902,PhysRevLett.89.247902,PhysRevA.71.022316,PhysRevLett.98.050502,PhysRevLett.114.200502,PhysRevA.95.062303,PhysRevB.95.125410}. A popular universal set consists of the controlled NOT gate~(CNOT) and either a collection of three fixed-angle single-qubit gates or another collection of four single-qubit gates~\cite{Williams2011}. In other words, to build a reliable quantum computer requires to prepare multiple gates of four or five types, each with an adequate accuracy. In 1995, Adriano Barenco introduced the following two-qubit quantum gate~\cite{Barenco1995},
\begin{eqnarray}
 \mathbb{B} &=& \left(\begin{array}{cccc}
    1 &0 &0 &0 \\
    0& 1 &0 &0\\
    0& 0& e^{i\alpha}\cos\theta & -ie^{i(\alpha-\phi)}\sin\theta \\
    0& 0 & -ie^{i(\alpha+\phi)}\sin\theta& e^{i\alpha}\cos\theta
  \end{array}
  \right),\label{BarencoGate}
\end{eqnarray}
which by itself constitutes a universal set, where $\alpha,\theta$, and $\phi$ are fixed irrational multiples of $\pi$ and of each other. Another universal gate similar to $\mathbb{B}$ could be found in Ref.~\cite{Sleator1995}. According to~\cite{Barenco1995}, being able to accurately realize $\mathbb{B}$ is sufficient for the construction of a quantum computation network. Since it is challenging to experimentally realize all quantum gates in a universal set with high accuracy, it seems a more attractive route to build a quantum computer by designing only one gate such as $\mathbb{B}$, compared with the strategy of designing several single-qubit gates and CNOT.

Although a single two-qubit gate as a universal set was proposed more than two decades ago~\cite{Barenco1995,Sleator1995}, its implementation remains an outstanding challenge. For the case of Barenco gate, it is possibly due to that $\mathbb{B}$ has three angles $\{\alpha,\theta,\phi\}$ although operating on two qubits. Thus designing the CNOT gate~(which is usually more challenging than realizing single-qubit gates) is the first choice~\cite{PhysRevLett.75.4714} with systems such as single photons~\cite{Pryde2003}, electrons in silicons~\cite{Veldhorst2015}, superconducting circuits~\cite{PhysRevLett.109.060501,Barends2014}, atomic ions~\cite{Ballance2016}, and neutral Rydberg atoms~\cite{Isenhower2010,Maller2015}.

\begin{figure}
\includegraphics[width=3.3in]
{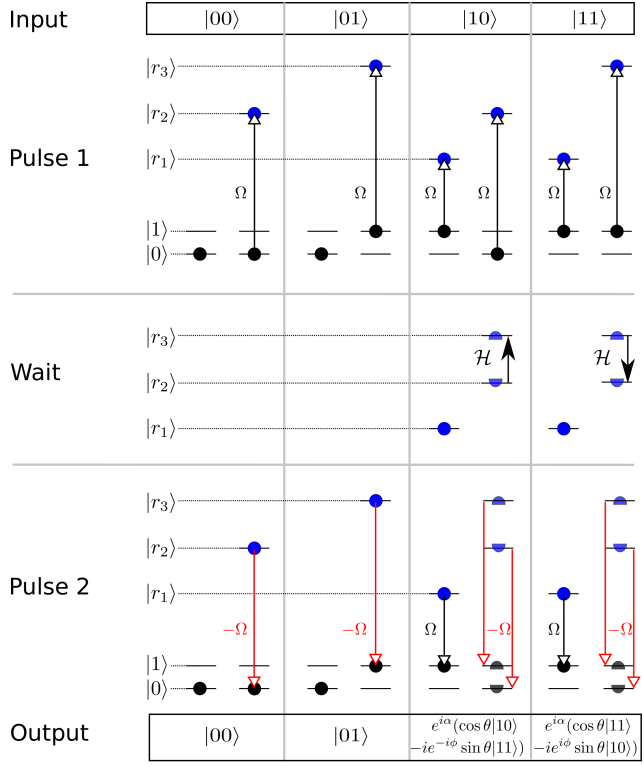}
 \caption{Protocol I for the Barenco gate. The first pulse maps qubit states to excited states, then the wait period allows the non-collinear interaction to induce state transformation that is essential for the gate, and the second pulse maps the excited states back to qubit states. There is a $\pi$ phase difference in the external fields between the first and second pulses upon the target qubit. \label{figGate} }
\end{figure}

Here we propose two protocols~(Protocol I and II) for Barenco gates when there is a non-collinear interaction between states that can be excited from the qubit states. Protocol I consists of two $\pi$ pulses and one wait period, illustrated in Fig.~\ref{figGate}, where $\phi$ is tunable by adjusting phases of external control, and $\alpha$ and $\theta$ change linearly with the wait duration in different ways. Protocol II consists of six $\pi$ pulses and two wait periods, where $\alpha$ changes linearly with the wait durations, while the other two variables depend on $\alpha$ non-linearly. Protocol I can lead to CNOT and Controlled-Y gates, and can be easily tuned to the parameter regime of $\{\alpha,\phi\}=\{1/4,1/2\}\pi$ and $\theta$ being an irrational multiple of $\pi$, where $\mathbb{B}$ becomes 
\begin{eqnarray}
 \mathbb{B}_{1} &=& \left(\begin{array}{cccc}
    1 &0 &0 &0 \\
    0& 1 &0 &0\\
    0& 0& e^{i\pi/4}\cos\theta & -ie^{-i\pi/4}\sin\theta \\
    0& 0 & ie^{-i\pi/4}\sin\theta&e^{i\pi/4}\cos\theta
  \end{array}
  \right),\label{BarencoGate02}
\end{eqnarray}
which, being similar to the universal gate introduced in~\cite{Sleator1995}, also constitutes a universal set by itself~\cite{Barenco1995}. Protocol II can also realize Eq.~(\ref{BarencoGate02}) if specific interactions are available.

Below, we detail the sequences for the two gate protocols, and analyze experimental prospects of realizing $\mathbb{B}$ by using van der Waals interaction~(vdWI) of Rydberg atoms~\cite{Gallagh2005}. Before proceeding, we introduce a generic method to construct a non-collinear interaction that is essential to our protocols.


\begin{figure}
\includegraphics[width=3.3in]
{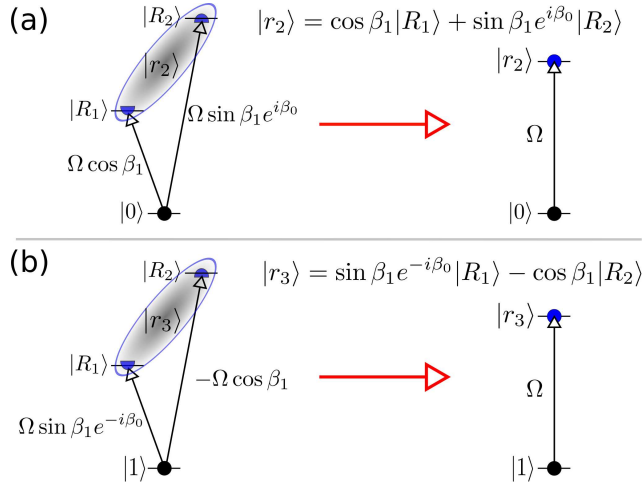}
 \caption{Excitation of the superposition states in Eq.~(\ref{defineRtor}) during pulse-1 of Protocol I. In (a) and (b), different superpositions in $|r_2\rangle$ and $|r_3\rangle$ follow after the different ratios of the two Rabi frequencies upon the energy eigenstates $|R_1\rangle$ and  $|R_2\rangle$. $\beta_1$ is tunable by adjusting the ratio between the strengths of the laser fields upon the two Rydberg eigenstates. \label{figBasis} }
\end{figure}

\section{Tunable non-collinear interaction}
  Our protocols are based on a non-collinear interaction between states excited from the qubit states, which we introduce here. We denote the basis states of Eq.~(\ref{BarencoGate}) by $\{|00\rangle,|01\rangle,|10\rangle,|11\rangle\}$, where $|\mu\nu\rangle\equiv |\mu\rangle_{\text{c}}\otimes|\nu\rangle_{\text{t}}$ is a two-qubit product state, with $|\mu(\nu)\rangle=|0\rangle$ or $|1\rangle$ being a ground state of a quantum system, and the subscripts c and t denote control and target, respectively. We suppose that no interaction exists in the four computational basis states of the gate. Among the four single-qubit states of the control and target, three can be connected to other states during the gate sequence,
\begin{eqnarray}
 |1\rangle_{\text{c}}\leftrightarrow  |r_1\rangle_{\text{c}}  ,~ |0\rangle_{\text{t}}\leftrightarrow|r_{2(3)}\rangle_{\text{t}},~ |1\rangle_{\text{t}}\leftrightarrow|r_{3(2)}\rangle_{\text{t}},\nonumber
\end{eqnarray}
where the kets right to ``$\leftrightarrow$'' are either excited states, or other ground states that can be connected with the qubit states $|0(1)\rangle$ via external control. In our gate sequence, the input states $|10\rangle$ and $|11\rangle$ can be excited to two states, $|r_1r_2\rangle$ and $|r_1r_3\rangle$, in which the following engineered two-body interaction arises,
\begin{eqnarray}
\mathcal{H} &=& \left(\begin{array}{ll}
    V_1 &V_{\text{e}}e^{-i\beta_0} \\
    V_{\text{e}}e^{i\beta_0}  & V_2
  \end{array}
  \right)  ,\label{vdW01}
\end{eqnarray}
which is written with the ordered basis $\{|r_1r_2\rangle,|r_1r_3\rangle\}$, where $\beta_0,V_1, V_2$ and $V_{\text{e}}$ are real variables. The off-diagonal interaction in Eq.~(\ref{vdW01}) is essential for our method. 

The form of Eq.~(\ref{vdW01}) is a little unusual compared with a more familiar two-body interaction of the following form,
\begin{eqnarray}
  \mathcal{H}_0 &=&  \sum_{j,k=0}^2b_{jk}|R_jR_k\rangle\langle R_jR_k| ,\label{generalV}
  \label{generalV}
\end{eqnarray}
where the energy eigenstates $|R_{0}\rangle,~|R_{1}\rangle$, and $|R_{2}\rangle$ are orthogonal to each other, and $b_{jk}$ is a blockade energy shift. Equation~(\ref{generalV}) can be found in various systems suitable for quantum computing, including~(but not limited to) electrons in quantum dots~\cite{RevModPhys.79.1217}, superconducting circuits~\cite{You2005}, and neutral Rydberg atoms~\cite{Saffman2010}. Nevertheless, a non-collinear interaction in Eq.~(\ref{vdW01}) rarely appears in qubits for quantum information processing, although it can be found in collective excitations in condensed matter systems~\cite{Xu2009,PhysRevB.93.161404}.

To realize Eq.~(\ref{vdW01}) based on Eq.~(\ref{generalV}), we consider the following orthogonal states
\begin{eqnarray}
  |r_1\rangle &=& |R_0\rangle, ~|r_2\rangle = \cos\beta_1|R_1\rangle+\sin\beta_1e^{i\beta_0} |R_2\rangle,\nonumber\\
  |r_3\rangle &=& \sin\beta_1e^{-i\beta_0} |R_1\rangle-\cos\beta_1|R_2\rangle\label{defineRtor}
\end{eqnarray}
in a rotating frame: $\hat{H}\rightarrow e^{i\hat{R}t}\hat{H}e^{-i\hat{R}t} -\hat{R}$, where $\hat{R}=\sum_j E_j|j\rangle\langle j|$ sums over all involved atomic states $|j\rangle$. In this rotating frame, the states $|r_{2}\rangle$ and $|r_{3}\rangle$ become eigenstates with any mixing angle $\beta_1\in[0,\pi/2]$. Notice that the superposition states above can also be written as
\begin{eqnarray}
  |r_2\rangle &=&e^{i\beta_0}( \cos\beta_1|R_1\rangle+\sin\beta_1 |R_2\rangle),\nonumber\\
  |r_3\rangle &=& \sin\beta_1 |R_1\rangle-\cos\beta_1|R_2\rangle\nonumber
\end{eqnarray}
by redefining $|R_{1(2)}\rangle$,  since a relative phase between $|R_1\rangle$ and $|R_2\rangle$ in $|r_{2(3)}\rangle$ is trivial in our case. $|r_{2}\rangle$ and $|r_{3}\rangle$ can be prepared by simultaneously exciting the two excited states $|R_1\rangle$ and $|R_2\rangle$ from the qubit state, shown in Fig.~\ref{figBasis}. A specific angle $\beta_0$ in Eq.~(\ref{defineRtor}) is determined by adjusting the phase of the external field upon the component $|R_{1(2)}\rangle$ to be $\beta_0(\beta_0-\pi)$ in $|r_{2}\rangle$ relative with that in $|r_{3}\rangle$, as shown in Fig.~\ref{figBasis}. Take neutral atoms as an example, $|0(1)\rangle$ is a hyperfine ground state, and $|R_1\rangle$ and $|R_2\rangle$ can be Rydberg eigenstates. When two laser beams of different frequencies are simultaneously sent upon one atom of initial state $|0\rangle$, with one laser exciting $|R_1\rangle$ with a Rabi frequency $\Omega\cos\beta_1$, and the other one pumping $|R_2\rangle$ with a Rabi frequency $\Omega\sin\beta_1e^{i\beta_0}$, shown in Fig.~\ref{figBasis}(a), a rotation between $|0\rangle$ and $|r_2\rangle$ with Rabi frequency $\Omega$ is established. Similarly, setting the two Rabi frequencies upon $|R_1\rangle$ and $|R_2\rangle$ as $\Omega\sin\beta_1e^{-i\beta_0}$ and $-\Omega\cos\beta_1$ establishes the preparation of $|r_3\rangle$.

According to Eq.~(\ref{generalV}), the interaction between the two orthogonal states $|r_1r_2\rangle$ and $|r_1r_3\rangle$ can be represented by Eq.~(\ref{vdW01}), where
\begin{eqnarray}
  V_1&=& b_{01}\cos^2\beta_1 +b_{02} \sin^2\beta_1,\nonumber\\
  V_2&=& b_{01}\sin^2\beta_1 +b_{02} \cos^2\beta_1,\nonumber\\
  V_{\text{e}}&=& (b_{01}-b_{02}) \sin\beta_1 \cos\beta_1.\label{Vfromb}
\end{eqnarray}
As long as $b_{01}\neq b_{02}$, $V_{\text{e}}$ can be nonzero. Notice that in Eq.~(\ref{defineRtor}), we can instead use $|R_1\rangle$ or $|R_2\rangle$ as $|r_1\rangle$. However, when an external control can simultaneously influence both qubits, it is necessary to choose $|r_1\rangle$ to be orthogonal to $|r_{2(3)}\rangle$. The three interaction strengths $V_1,~V_2$, and $V_{\text{e}}$ can be tuned by varying $\beta_1$, and the ratio $b_{01}/b_{02}$ is adjustable via choosing different sets of states $\{|R_0\rangle,|R_1\rangle,|R_2\rangle\}$. 

Below we present two protocols based on Eq.~(\ref{vdW01}), where the three variables $\alpha,\theta$ and $\phi$ exhibit distinct tunabilities that can be beneficial for different purposes in quantum control.
\begin{figure}
\includegraphics[width=3.3in]
{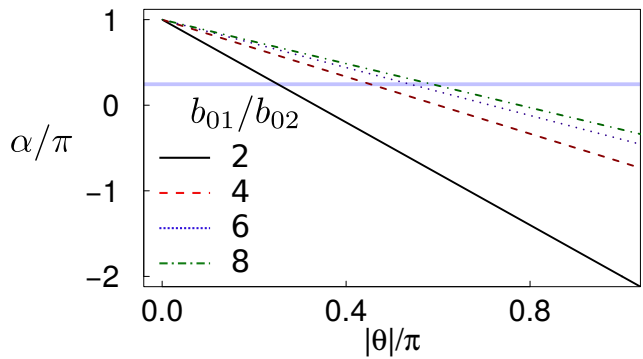}
 \caption{$\alpha$ as a function of $|\theta|$ in Protocol I for the $\mathbb{B}$ gate with several different values of $b_{01}/b_{02}$. $\theta$ is positive~(negative) if $b_{01}-b_{02}>0(<0)$. The other variable $\phi$ in the gate is freely tunable. The horizontal gray line locates $\alpha=\pi/4$, a value that satisfies the condition for another universal gate when $\phi=\pi/2$ and $b_{01}/b_{02}$ is irrational~\cite{Barenco1995}.  \label{figTunability01} }
\end{figure}

\section{Protocol I: A two-pulse sequence}
We first show a two-pulse sequence for $\mathbb{B}$ when $V_1=V_2$ and $\beta_0$ in Eq.~(\ref{vdW01}) is tunable. When $|R_{j}\rangle$ is a state of neutral atoms, where $j=0-2$, $\beta_0$ can be tuned by varying the relative phases among the external control fields.

As illustrated in Fig.~\ref{figGate}, Protocol I starts with a $\pi$ pulse of Rabi frequencies $\Omega$ upon the control and target qubit states $|1\rangle_{\text{c}},|0\rangle_{\text{t}}$, and $|1\rangle_{\text{t}}$,
\begin{eqnarray}
  \{|1\rangle_{\text{c }}, |0\rangle_{\text{t }}, |1\rangle_{\text{t }} \} &\mapsto & -i\{|r_1\rangle_{\text{c }}, |r_2\rangle_{\text{t }}, |r_3\rangle_{\text{t }}\}. \label{pi-01}
\end{eqnarray}
When $\Omega\gg \{V_1,V_2,V_{\text{e}}\}$, we have the following map,
\begin{eqnarray}
  \{|00\rangle,|01\rangle,|10\rangle, |11\rangle\}\mapsto-\{i|0r_2\rangle, i|0r_3\rangle,|r_1r_2\rangle,|r_1r_3\rangle\}.\nonumber
\end{eqnarray}
The process above is subject to a residue blockade effect which can be minimized by increasing $\Omega$ relative to $V_{1(2)}$ and $V_{\text{e}}$. 

Upon completion of pulse-$1$, a wait period of duration $T$ is allowed when the two-atom state evolves under the interaction in Eq.~(\ref{vdW01}),
\begin{eqnarray}
  |r_1r_2\rangle & \mapsto & \eta_1 e^{-iT\lambda_+ } |\lambda_+ \rangle + \eta_2e^{-iT\lambda_- +i\beta_0}|\lambda_-\rangle , \nonumber\\
  |r_1r_3\rangle & \mapsto & \eta_2e^{-iT\lambda_+-i\beta_0 } |\lambda_+ \rangle - \eta_1e^{-iT\lambda_- } |\lambda_-\rangle . \label{waitEvo}
\end{eqnarray}
Here
\begin{eqnarray}
\lambda_\pm &=& (V_1+V_2)/2 \pm \overline{V},\nonumber\\
  |\lambda_+\rangle &=& \eta_1|r_1r_2\rangle + \eta_2e^{i\beta_0}|r_1r_3\rangle , \nonumber\\
  |\lambda_-\rangle &=& \eta_2e^{-i\beta_0}|r_1r_2\rangle - \eta_1|r_1r_3\rangle,\nonumber
\end{eqnarray}
are the eigenvalues and normalized eigenvectors of Eq.~(\ref{vdW01}), where $\overline{V}  = \sqrt{V_{\text{e}}^2 +(V_1-V_2)^2/4 }$ and $\eta_1 :\eta_2 =V_{\text{e}} : (2\overline{V}+V_2-V_1) /2$. For the sake of convenience, a frequently appeared Planck constant is hidden.

Soon after the wait period, another set of external fields with strengths similar to those in the first pulse are applied, with a $\pi$ phase shift in the control fields upon the target qubit. The Rabi frequency on the control is still $\Omega$, but those on the target become $-\Omega$, so as to induce the following map,
\begin{eqnarray}
\{|r_1\rangle_{\text{c }}, |r_2\rangle_{\text{t }}, |r_3\rangle_{\text{t }}\} &\mapsto & i  \{-|1\rangle_{\text{c }}, |0\rangle_{\text{t }}, |1\rangle_{\text{t }} \} ,\label{pi-02}
\end{eqnarray}
which differs from Eq.~(\ref{pi-01}) in that the phase change to a state of the target qubit is $\mp\pi/2$ in Eq.~(\ref{pi-01})[(\ref{pi-02})].

As can be easily verified, the state evolution from the input to output under the condition of $V_1=V_2$~[or equivalently, $\beta_1=\pm \pi/4$ in Eq.~(\ref{Vfromb})] is, 
\begin{eqnarray}
  |00\rangle&\mapsto& |00\rangle,\nonumber\\
  |01\rangle&\mapsto& |01\rangle ,\nonumber\\
  |10\rangle&\mapsto& e^{i\alpha} (-i e^{-i\phi}\sin\theta|11\rangle + \cos\theta |10\rangle ) ,\nonumber\\
  |11\rangle&\mapsto& e^{i\alpha} ( \cos\theta|11\rangle-i e^{i\phi}\sin\theta|10\rangle  ).\label{BZgate2}
\end{eqnarray}
Here 
\begin{eqnarray}
 \alpha &=&\pi-V_1T=\pi-(b_{01}+b_{02})T/2,\nonumber\\
  \theta &=& V_{\text{e}}T=\pm|b_{01}-b_{02}|T/2,  \nonumber\\
  \phi &=& -\beta_0,\label{anglesI}
\end{eqnarray}
where $+(-)$ in $\theta$ applies for a positive~(negative) $V_{\text{e}}$. Equation~(\ref{BZgate2}) is exactly the gate $\mathbb{B}$ in Eq.~(\ref{BarencoGate}). Here $\phi$ is determined by phases of external control fields, thus is tunable and independent of $\{\alpha,\theta\}$. $\alpha$ and $\theta$ depend on the wait duration and the interaction strengths $b_{01}$ and $b_{02}$, and have a linear relation with each other,
\begin{eqnarray}
 \alpha &=&\pi-\frac{b_{01}+b_{02}}{b_{01}-b_{02}}\theta.\label{alphaI}
\end{eqnarray}
When $b_{01}-b_{02}>0$, $\alpha$ is shown in Fig.~\ref{figTunability01} as a function of $\theta$ with several sets of $b_{01}/b_{02}$. As proved in Ref.~\cite{Barenco1995}, Eq.~(\ref{BZgate2}) is a useful Barenco gate when $\alpha,\theta$, and $\phi$ are irrational multiples of $\pi$ and of each other. In the above protocol, $\theta$ can be tuned by choosing an appropriate $T$ to be any irrational multiple of $\pi$, and simultaneously $\alpha$ is an irrational multiple of $\pi$ at least for a rational $b_{01}/b_{02}$, according to Eq.~(\ref{alphaI}). Finally, $\phi$ can be tuned to any value via varying relative phases of external fields. So, there should be infinite sets of $\alpha,\theta$, and $\phi$ that are irrational multiples of $\pi$ and of each other.

From Ref.~\cite{Barenco1995}, when $\{\alpha,\phi\}=\{1/4,1/2\}\pi$ and $\theta$ is an irrational multiple of $\pi$, the gate in Eq.~(\ref{BZgate2}) also constitutes a universal set. Equation~(\ref{alphaI}) indicates that $\theta$ is an irrational multiple of $\pi$ when $b_{01}/b_{02}$ is irrational and $\alpha=\pi/4$. Since the condition of $\alpha=\pi/4$ is readily achievable as indicated by the horizontal line in Fig.~\ref{figTunability01}, and $b_{01}/b_{02}$ in a real system can be an irrational number infinitely near to an rational $b_{01}/b_{02}$ such as those in Fig.~\ref{figTunability01}, Protocol I can also construct the universal gate in Eq.~(\ref{BarencoGate02}) besides the Barenco gate.

\begin{figure}
\includegraphics[width=3.3in]
{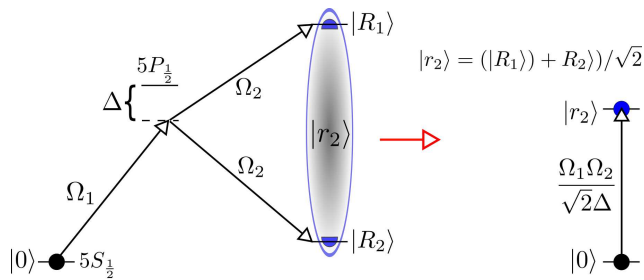}
 \caption{Scheme of constructing $|r_2\rangle$ so that $b_{02}=0$ in Eq.~(\ref{Vfromb}) for realizing CNOT and Control-Y gates. Here $\Delta$ is a detuning that is large compared with $\Omega_1$ and $\Omega_2$. A similar configuration for $|r_3\rangle$ is realized when the two Rabi frequencies upon $|R_1\rangle$ and $|R_2\rangle$ are $\Omega_2$ and $-\Omega_2$. \label{R2ground} }
\end{figure}
\subsection{CNOT and Controlled-Y gates}\label{CXCY}
Below, we show that Protocol I can also lead to CNOT and Control-Y gates. Although such gates can not constitute a universal set unless single-qubit rotations are brought in, realization of an ``over-complete'' family of universal gates may be helpful to construct a quantum computing circuit~\cite{Williams2011}. Meanwhile, though a single quantum gate as a universal set has certain advantages such as lower overhead in calibration, it can be less efficient for certain quantum computation algorithms. In principle, single-qubit rotations together with any two-qubit entangling gate can build up a quantum computing network, and an entangling gate which is less challenging to realize is often the favorite for current interest, which is why the CNOT gate has received widespread attention.
 
Protocol I can easily lead to a CNOT gate. As seen from Eq.~(\ref{anglesI}), when $b_{01}>0$ and $b_{02}=0$, a wait time $T=\pi/b_{01}$ in Protocol I leads to $\alpha=\theta=\pi/2$. Setting $\beta_0=0$ ensures $\phi=0$, then the gate transformation of Protocol I in Eq.~(\ref{BZgate2}) becomes
\begin{eqnarray}
  |00\rangle&\mapsto& |00\rangle,~~  |01\rangle\mapsto |01\rangle ,\nonumber\\
  |10\rangle&\mapsto&  |11\rangle  , ~~ |11\rangle\mapsto|10\rangle
\end{eqnarray}
as in a CNOT gate. The requirement in this CNOT gate can be easily set for neutral atoms. First, the condition $b_{01}>0$ is fulfilled by choosing both $|r_1\rangle$ and $|R_1\rangle$ in Eq.~(\ref{defineRtor}) from a common high-lying s-orbital Rydberg state $|R_0\rangle$. Take $|r_1\rangle=|R_1\rangle=|96s_{1/2},m_J=1/2,m_I=3/2$ as an example, the calculated~\cite{Shi2014} interaction coefficient $b_{01}=36\times2\pi$~THz~$(\mu m/l)^6$ is about $0.6\times2\pi$~MHz even when the two qubits are separated by a large distance of $l=20\mu m$.

The other requirement for realizing a CNOT gate, $b_{02}=0$, is achievable by choosing $|R_2\rangle$ from a ground state, since the interaction between Rydberg and ground states can be neglected. Specifically, if $|0(1)\rangle = |5s_{1/2},F=1(2),m_F=-1\rangle$, one can choose $|R_2\rangle$ as $|5s_{1/2},F=1,m_F=1\rangle$, a state that can be reached from both $|0\rangle$ and $|1\rangle$ by using detuned circularly polarized laser fields upon an intermediate state $|5p_{1/2},F=1,m_F=0\rangle$. Take $|0\rangle\rightarrow|r_2\rangle$ as an example, its excitation is shown in Fig.~\ref{R2ground}, with its effective Rabi frequency $\Omega$ given by $\Omega_1\Omega_2/\sqrt2\Delta$~\cite{Shi2014}. Because circularly polarized laser fields induce transitions between two levels differing in hyperfine quantum numbers by $\Delta m_F=\pm1$, the laser connecting $|5p_{1/2},F=1,m_F=0\rangle$ and $|R_2\rangle$ in Fig.~\ref{R2ground} may also couple $|0(1)\rangle$ leftward to a level with $m_F=-2$. Such a coupling, however, is negligible via the selected $|5p_{1/2},F=1\rangle$ manifold because it does not host a state with $m_F=-2$. Similarly, the laser addressing $|0(1)\rangle\leftrightarrow |5p_{1/2},F=1,m_F=0\rangle$ in Fig.~\ref{R2ground} can not couple $|R_2\rangle$ with $|5p_{1/2},F=1\rangle$ since it does not host a state with $m_F=2$. Alternatively, level shifting by a strong enough external magnetic field can be applied to avoid population leakage if we use a coupling scheme different from that in Fig.~\ref{R2ground}.

Protocol I can also realize the Controlled-Y gate. Still, we use a similar setting described above to reach the domain of $b_{01}>0$ and $b_{02}=0$ so that $\alpha=\theta=\pi/2$ can be achieved by choosing a wait duration of $T=\pi/b_{01}$. Different from the CNOT gate above, here the phase of laser field shall render $\beta_0=-\pi/2$ so that $\phi=\pi/2$. Then, Eq.~(\ref{BZgate2}) shows that the gate maps the states according to
\begin{eqnarray}
  |10\rangle&\mapsto&  -i|11\rangle  , ~~ |11\rangle\mapsto  i |10\rangle,
\end{eqnarray}
while the other two input states $|00\rangle$ and $|01\rangle$ are not affected, realizing the Controlled-Y gate.

\begin{figure}
\includegraphics[width=3.3in]
{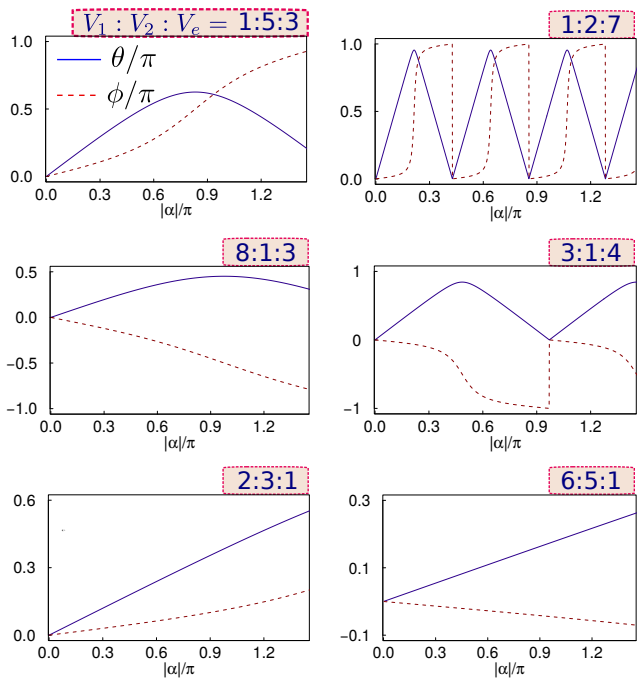}
 \caption{The two variables $\theta$ and $\phi$ as a function of $\alpha$ in Protocol II for the $\mathbb{B}$ gate. Here we choose six different sets of $(V_1:V_2 :V_{\text{e}})$, shown on top of each subfigures.   \label{figTunability} }
\end{figure}

\section{Protocol II: A six-pulse sequence}
Below, we describe Protocol II where each of the three angles $\{\alpha,\theta,\phi\}$ depends on model parameters in a different manner compared with that in Protocol I. The six-pulse protocol below is based on $V_1\neq V_2$ and $\beta_0=0$ in Eq.~(\ref{vdW01}). A six-pulse sequence is chosen so as to show details, although we can also adopt a four-pulse sequence because the first~(last) two pulses can occur simultaneously. For the sake of convenience, we use pulse-$k$ to denote the $k$th pulse, where $k=1-6$.

Pulse-$1$ is a $\pi$ pulse upon the state $|1\rangle_{\text{c}}$ of the control qubit, so that
\begin{eqnarray}
 \{|10\rangle, |11\rangle\}\mapsto-i\{|r_10\rangle, |r_11\rangle\},\nonumber
\end{eqnarray}
while the other two states $|00\rangle$ and $|01\rangle$ stay intact.

Pulse-$2$ is a simultaneous $\pi$ pulse upon qubit states $|0\rangle_{\text{t}},|1\rangle_{\text{t}}$. When $\Omega\gg \{V_1,V_2,V_{\text{e}}\}$, we have the following map,
\begin{eqnarray}
  \{|00\rangle,|01\rangle,-i|r_10\rangle,-i|r_11\rangle\}&\mapsto&-\{i|0r_2\rangle, i|0r_3\rangle,|r_1r_2\rangle,\nonumber\\
  &&~~~~|r_1r_3\rangle\}.\nonumber
\end{eqnarray}

Upon completion of pulse-$2$, a wait period of duration $T$ is allowed when the two-qubit state evolves under the interaction in Eq.~(\ref{vdW01}), where the state evolution is identical to that in Eq.~(\ref{waitEvo}), with $\beta_0=0$.

Pulse-$3$ maps the states of the target qubit to ground states, i.e., inverse to pulse-$2$,
\begin{eqnarray}
\{ |r_2\rangle_{\text{t }}, |r_3\rangle_{\text{t }}\} &\mapsto & -i \{ |0\rangle_{\text{t }}, |1\rangle_{\text{t }} \}.\nonumber
\end{eqnarray}

Pulse-$4$ is also a $\pi$ pulse but maps the states of the target qubit to excited states in a different manner compared with pulse-$2$,
\begin{eqnarray}
  \{ |0\rangle_{\text{t }}, |1\rangle_{\text{t }} \} &\mapsto &-i \{ |r_3\rangle_{\text{t }}, |r_2\rangle_{\text{t }}\}. \label{Pulse4}
\end{eqnarray}

Upon completion of pulse-$4$, we again allow a wait period of duration $T$, so that the four input states $\{|00\rangle,|01\rangle,|10\rangle,|11\rangle\}$ evolve respectively to, $\{i|0r_3\rangle,i|0r_2\rangle,\chi_1|r_1r_2\rangle+\chi_2|r_1r_3\rangle,\chi_3|r_1r_2\rangle+\chi_4|r_1r_3\rangle  \}$, where
\begin{eqnarray}
  (\chi_1,\chi_2,\chi_3,\chi_4)&=&e^{i\alpha}\cos\theta (i e^{-i\phi}\tan\theta,-1,-1,i e^{i\phi}\tan\theta) ,\nonumber
\end{eqnarray}
and
\begin{eqnarray}
  \alpha&=&-T(V_1+V_2),\nonumber\\
  \sin\theta &=& 2(\eta_1\eta_2)\big\{(\eta_1^2-\eta_2^2)^2[\cos( 2T\overline{V})-1]^2\nonumber\\
  && ~+\sin^2( 2T\overline{V}) \big\}^{1/2},\nonumber\\
  \cos\theta &=& 1+4\eta_1^2\eta_2^2[\cos( 2T\overline{V})-1],\nonumber\\
  \sin\phi &=& 2(\eta_1\eta_2)(\eta_1^2-\eta_2^2)[\cos( 2T\overline{V})-1] / |\sin\theta|,\nonumber\\
  \cos\phi &=&2(\eta_1\eta_2) \sin( 2T\overline{V})/ |\sin\theta|. \label{anglesII}
\end{eqnarray}

Pulse-$5$ is identical to pulse-$4$ so as to induce a state transformation inverse to Eq.~(\ref{Pulse4}), and finally, pulse-$6$ is inverse to pulse-$1$. As a consequence, the overall effect on the four input states is described by Eq.~(\ref{BZgate2}), which can be represented in a matrix form of Eq.~(\ref{BarencoGate}).

Similar to Protocol I, in Protocol II there are many cases where $\alpha,\theta$, and $\phi$ are irrational multiples of $\pi$ and of each other. $\alpha,\theta$, and $\phi$ are functions of the three variables $V_1T, V_2T$ and $V_{\text{e}}T$. For each set $\{V_1, V_2, V_{\text{e}}\}$, when $T$ changes, $\alpha$ changes linearly, but $\theta$ and $\phi$ evolve non-linearly. By choosing several sets of $(V_1:V_2:V_{\text{e}})$, we show in Fig.~\ref{figTunability} that $\theta$ and $\phi$ evolve in different ways, indicating the existence of many choices of $\alpha,\theta$, and $\phi$ that are irrational multiples of $\pi$ and of each other.

Protocol II also allows realization of $\mathbb{B}_{1}$ in Eq.~(\ref{BarencoGate02}). The condition of $\{\alpha,\phi\}=\{1/4,1/2\}\pi$ can be satisfied with $T=\pi/2\overline{V}$ when $V_1+V_2=-\overline{V}/2$, where the latter condition requires $-b_{01}/b_{02}=5/3$ or $3/5$. If one realizes $\mathbb{B}_{1}$ by using Rydberg interaction of neutral atoms, the desired $b_{01}/b_{02}$ can be reached either by using external fields to tune the energy gaps between appropriate Rydberg levels, or by introducing another independent variable $\beta_2$ when choosing a superposition state for $|R_2\rangle$ in Eq.~(\ref{defineRtor}),
\begin{eqnarray}
  |R_2\rangle \rightarrow \cos\beta_2|R_2\rangle+\sin\beta_2 |R_3\rangle,\label{beta2}
\end{eqnarray}
so that the parameter $b_{02}$ in Eq.~(\ref{Vfromb}) becomes tunable by varying $\beta_2$,
\begin{eqnarray}
b_{02} &\rightarrow& b_{02}\cos^2\beta_2 +b_{03} \sin^2\beta_2.\nonumber
\end{eqnarray}
 When $b_{02}<b_{03}$, the scheme above transfers the former $b_{02}$ to a new one tunable in the interval $[b_{02},b_{03}]$. In case Eq.~(\ref{beta2}) is adopted, it is necessary to use microwave fields that are strong enough to suppress the transition from $\cos\beta_2|R_2\rangle+\sin\beta_2 |R_3\rangle$ to $(\cos\beta_2|R_3\rangle-\sin\beta_2 |R_2\rangle)$, as detailed in~\cite{Shi2017pra}.

\section{Realization with neutral atoms}
We turn to analyze the feasibility of realizing the protocols above with two neutral $^{87}$Rb atoms. As for the qubit states, one can choose from the hyperfine ground states $|0(1)\rangle = |5s_{1/2},F=1(2),m_F=0\rangle$ 
for both the control and target~\cite{Isenhower2010}, where $\{F,m_F\}$ constitute the hyperfine notation of the ground states. For Rydberg states introduced below, however, we apply the fine structure notation according to the spectroscopic resolution achieved in experiments. 

The interaction defined in Eqs.~(\ref{vdW01}) and~(\ref{Vfromb}) can be from vdWI between neutral atoms, which we briefly introduce here. When each of two nearby neutral atoms is in a Rydberg state $|R_0\rangle$, a strong interaction between the electric dipole moments of the two atoms can arise. When this interaction energy is much smaller than the energy gaps to other nearby two-atom Rydberg states, it gives an overall energy shift $b_{00}$~[see Eq.~(\ref{generalV})] to the two-atom Rydberg state~\cite{Gallagh2005}. Such vdWI can be several tens of megahertz, which is strong enough to induce fast quantum dynamics on the single-quantum level~\cite{Dudin2012}. This means that when one atom is already in a Rydberg state $|R_0\rangle$, it is difficult to use resonant fields to excite a nearby atom to $|R_0\rangle$ unless the applied field is very strong. Based on this blockade interaction, Ref.~\cite{PhysRevLett.85.2208} proposed two-qubit controlled-phase gates, one of which was experimentally demonstrated years ago~\cite{Isenhower2010}. Since then, there have been various proposals of two-qubit controlled-phase~(or CNOT) gates based on Rydberg blockade~\cite{Saffman2016}. Instead of focusing on the blockade mechanism, this work relies on the exchange interaction of Rydberg atoms, which was much less explored either experimentally~\cite{Thompson2017} or theoretically~\cite{Shi2014,Beterov2016,Petrosyan2017} for the purpose of quantum information processing. The exchange interactions used in~\cite{Thompson2017,Shi2014,Beterov2016,Petrosyan2017}, however, are limited to the type where the states of both atoms simultaneously change. In contrast, the exchange process of Eq.~(\ref{vdW01}) in this work only changes the state of the target qubit.

\begin{figure}\centering
\includegraphics[width=2.3in]
{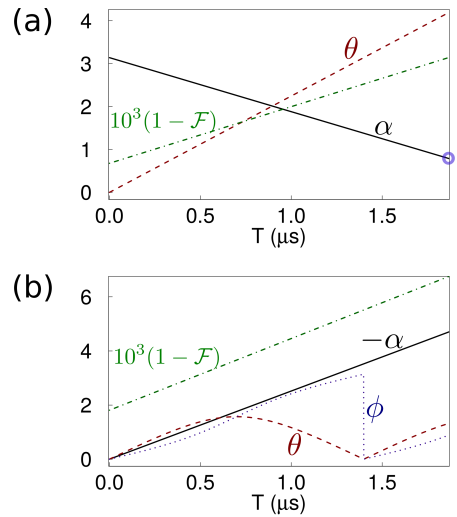}
 \caption{The solid, dashed, dotted, and dash-dotted curves respectively show the three variables $\alpha$, $\theta$, and $\phi$, and the total error~(scaled up by $10^3$) of the intrinsic gate fidelity as a function of $T$ in the $\mathbb{B}$ gate protocols realized with two neutral $^{87}$Rb atoms. (a) and (b) are for Protocols I and II, respectively, and the circle on the curve of $\alpha$ in (a) locates the value of $\alpha=\pi/4$. \label{figError} }
\end{figure}

We illustrate the performance of the gate by choosing
\begin{eqnarray}
 |R_0\rangle& =&|R_1\rangle= |n_1s_{1/2},m_J=1/2,m_I=3/2\rangle,\nonumber\\
|R_2\rangle &=& |n_2s_{1/2},m_J=1/2,m_I=3/2\rangle \label{defineRtorRydberg}
\end{eqnarray}
for the construction of $|r_j\rangle$ in Eq.~(\ref{defineRtor}), where $n_1$ and $n_2$ are two different large principal quantum numbers, and $m_J$~($m_I$) denotes the electric~(nuclear) spin projection on the quantization axis. Preparation of a superposition Rydberg state $|r_{2(3)}\rangle$ can be performed via two-photon excitation in the ``V'' or ``Y'' configuration, as shown in Fig.~\ref{figBasis} and detailed in~\cite{ShiJPB2016,Shi2017pra}, although here we need not to stabilize the superposition by extra external fields unless Eq.~(\ref{beta2}) is used.

\subsection{Intrinsic fidelity}\label{sectionVA}
Regarding the gate fidelity $\mathcal{F}$~\cite{Poyatos1997}, the Rydberg state decay, blockade errors from incomplete rotations, and population leakage to levels other than the qubit states render imperfect gate operation characterized by a fidelity error $1-\mathcal{F}$. These errors can be estimated from analytical approximations~\cite{Saffman2005}. For example, decay-induced error can be estimated from the fact that the population loss due to Rydberg state decay is proportional to the time of populating a Rydberg state. We estimate $1-\mathcal{F}$ by choosing $\{n_1,n_2\}=\{96,102\}$, a two-atom distance of $20\mu$m, a Rabi frequency $\Omega/2\pi=30$~MHz, and an environment temperature of $4$~K. The vdWI for this choice can be found in Appendix~\ref{appA}. We let $\beta_1$ be $\pi/4$ and $3\pi/8$ for Protocols I and II, respectively, since $\beta_1$ should~(should not) be $\pm\pi/4$ for Protocol I~(II): If $|\beta_1|=\pi/4$ in Protocol II, $\phi$ becomes $0,\pm\pi,\cdots$. We show the variables $\alpha,\theta$, and $\phi$ and the fidelity error rescaled by $10^3$ in Fig.~\ref{figError} for both Protocols I and II, according to Eqs.~(\ref{anglesI}) and~(\ref{anglesII}) and the estimates in Appendix~\ref{appB}. $\phi$ in Protocol I is not shown in Fig.~\ref{figError}(a) because it is determined by phases of external fields. If we instead assume a temperature of $300$~K, a larger error from Rydberg state decay occurs and the fidelity error $1-\mathcal{F}$ in Fig.~\ref{figError} increases to be in the interval $[0.7,~5.7]([2.0,~12])\times10^{-3}$ for Protocol I~(II). Here two other error-causing factors have been ignored. First, we find that errors from the force between the two atoms when both of them are in Rydberg states can be ignored, as shown in Appendix~\ref{appB}. Second, an extra error $\overline{E}_{\mathcal{L}}$ from the position fluctuation of the atoms can be neglected, too. To show the smallness of $\overline{E}_{\mathcal{L}}$, we assume optical tweezer traps created by single laser beams with wavelength $\lambda=1.1\mu$m and waist $w=3\mu$m. If the atoms are not cooled to motional ground states before the gate sequence, numerical calculation as in~\cite{Shi2017pra,Shi2017} shows that $\overline{E}_{\mathcal{L}}\in[1.4,~52]([1.0,~49])\times10^{-4}$ for Protocol I~(II) when the effective atomic temperature $T_a\in[10,200]\mu$K and the trap depth is $U=20$~mK. We also considered a similar set-up analyzed in Ref.~\cite{Shi2017} when atoms are cooled to motional ground states, and found that $\overline{E}_{\mathcal{L}}<7\times10^{-5}$ for both Protocol I and II when $U>1\mu$K. So, The error caused by position fluctuation can be suppressed by sufficient cooling of the atoms.

The gate fidelity in most cases of Fig.~\ref{figError} is heavily hampered by decay probability of Rydberg states, which is mainly determined by wait duration $T$ and thus inversely proportional to the vdWI parameters $V_1,V_2$, and $V_{\text{e}}$~[see Eqs.~(\ref{anglesI}) and~(\ref{anglesII})]. Appendix~\ref{appA} shows that $V_1,V_2$, and $V_{\text{e}}$ of Fig.~\ref{figError} are smaller than $2\pi$~MHz, resulting in microsecond-scale gate times and significant Rydberg state decay. But if we use larger vdWI by decreasing qubit spacing, the calculated vdWI can approach the energy gap to nearby two-qubit Rydberg states and violates the picture of vdWI. However, it is possible to tackle this issue via pulse shaping, a technique useful in neutral atoms~\cite{Theis2016} as well as solid-state systems~\cite{PhysRevB.95.241307}.

In principle, the intrinsic gate fidelity error in Fig.~\ref{figError} can be significantly suppressed by recently proposed schemes. For example, the blockade error can be removed by exploring rational generalized Rabi frequencies in detuned Rabi transitions~\cite{Shi2017}, and population leakage can be reduced in the adiabatic regime~\cite{Petrosyan2017}. Our purpose here, however, is to provide simplest protocols for $\mathbb{B}$ so as to inspire further exploration of quantum information processing by using the interaction in Eq.~(\ref{vdW01}).

\subsection{Tunable operation modes}
Four parameters are tunable in the two protocols: the frequency, intensity and phase of laser fields of optical pulses, and the wait duration $T$ between the optical pulses. First of all, the interaction coefficients $b_{jk}$ can be adjusted by choosing different eigenstates in Eqs.~(\ref{generalV}) and~(\ref{defineRtorRydberg}) via using lasers of different frequencies. If the eigenstates $|R_{0}\rangle$ and $|R_{1}\rangle$ in Eq.~(\ref{generalV}) are Rydberg states, but $|R_{2}\rangle$ is a ground state, then we have the condition of $b_{02}=0$ for the realization of CNOT and Controlled-Y gates, as described in Sec.~\ref{CXCY}. Below, we discuss the tunability when the eigenstates $|R_{j(k)}\rangle$ in Eq.~(\ref{generalV}) are all Rydberg states.

For every set of $|R_{j(k)}\rangle$ in Eq.~(\ref{generalV}), the angle $\beta_1$ in Eq.~(\ref{defineRtor}) can be tuned by adjusting the intensity of laser fields. For example, when lasers in Fig.~\ref{figBasis} are set in a way that $\beta_1=k\pi/2$ with an integer $k$, the non-collinear coefficient $V_{\text{e}}$ disappears in Eq.~(\ref{vdW01}), which is a case neither Protocol I nor II lead to a Barenco gate. By continuously changing the magnitudes of the laser fields upon the two Rydberg states $|R_{1(2)}\rangle$ in Fig.~\ref{figBasis}(a)~[or (b)],  the mixing angle $\beta_1$ in the definition of $|r_2\rangle$~[or $|r_3\rangle$] can be continuously changed. When $\beta_1=\pi/4$, $V_1=V_2$ is achieved and Protocol I can be realized.

The ratio between $V_1,~V_2$, and $V_{\text{e}}$ can be tuned by changing the mixing angle $\beta_1$ in Protocol II, so that the scaling of the angles $\alpha$, $\theta$, and $\phi$ can behave in distinct ways, shown in Fig.~\ref{figTunability}. For instance, in case $b_{01} \gg b_{02}>0 $, we have $(V_1:V_2 :V_{\text{e}})\approx (\cos^2\beta_1:\sin^2\beta_1:\sin\beta_1\cos\beta_1 )$, which includes at least two cases: $V_1>V_{\text{e}}>V_2$ and $V_2>V_{\text{e}}>V_1$. Similarly, if $b_{01} > -b_{02}>0 $, we can realize another pair of cases: $V_{\text{e}}>V_1>V_2$ or $V_{\text{e}}>V_2>V_1$. When $b_{01} \sim b_{02}>0$, adjusting $\beta_1$ can lead to $V_1>V_2>V_{\text{e}}$ and $V_2>V_1>V_{\text{e}}$, covering all cases in Fig.~\ref{figTunability}.

Furthermore, the angle $\beta_0$ in Eq.~(\ref{defineRtor}) can be tuned by adjusting the relative phases of laser fields upon the two Rydberg states $|R_{1(2)}\rangle$ in Figs.~\ref{figBasis}(a) and~\ref{figBasis}(b). This is crucial for Protocol I: although the angles $\alpha$ and $\theta$ in Eq.~(\ref{BarencoGate}) are determined by the wait duration $T$, the angle $\phi$ is determined by phases of the laser fields, shown in Eq.~(\ref{anglesI}).

Finally, the wait duration $T$ in both protocols can be varied, so that the angles $\alpha$ and $\theta$ in Protocol I, and all the three angles $\alpha$, $\theta$, and $\phi$ in Protocol II can be tuned.

In summary, when the frequency, intensity, and phase of laser fields, and the wait duration between laser pulses are tuned, various sets of the three angles in the Barenco gate can be realized.

\subsection{Experimental prospects}
High-fidelity realization of the above protocols depends on availability of strong enough pulsed lasers, since the blockade error can be suppressed only when $\Omega\gg \{V_1,V_2,V_{\text{e}}\}$. Take the analysis leading to Fig.~\ref{figError} as an example, where $ V_1,V_2,V_{\text{e}}\lessapprox0.36\times2\pi$~MHz, a two-photon Rabi frequency larger than $ 3\times2\pi$~MHz in $\pi$ pulses of the protocols is preferable. This is in principle possible: Refs.~\cite{PhysRevLett.107.243001,PhysRevLett.114.203002} reported coherent GHz-rate Rabi oscillations between ground and $nS_{1/2}$ Rydberg states with $n\geq30$ via laser pulses of nanosecond durations upon a rubidium~(cesium) vapor. For single cold atoms, $\pi$ pulses with Rabi frequency of $7\times2\pi$~MHz between ground and $58d_{3/2}$ states of $^{87}$Rb were used in Ref.~\cite{Gaetan2009}. Since the Rydberg states in our example are relatively high, we assume that a Rabi frequency of $\Omega/2\pi=5$~MHz can be easily realized for a conservative estimate. With such an $\Omega$, we find that the curve of $1-\mathcal{F}$ in Fig.~\ref{figError}(a)~[(b)] rises by $6\times10^{-3}$~[$32\times10^{-3}$].

Stable laser sources are also required to achieve the predicted gate performance in Fig.~\ref{figError}. Our protocols require the establishment of the superposition states defined in Eq.~(\ref{defineRtor}), whose preparation depends on correctly setting the magnitude and phase of the laser fields in Fig.~\ref{figBasis}, provided that the laser frequency is stable enough~\cite{DeHond2017,Legaie2017}. We take Protocol I as an example to estimate the necessary precision for the parameters of the laser fields to reach the gate fidelity in Fig.~\ref{figError}. If the phase from dipole matrix element is fixed to be zero, then $\beta_0$ in Eq.~(\ref{vdW01}) is determined by phases of laser fields. Suppose the relative fluctuations of the laser phase and the laser Rabi frequency are bounded by $\varsigma_1$ and $|\delta\Omega_k/\Omega_k|\leq\varsigma_2$, then the phase term $\beta_0$ and interaction coefficients $V_1,~V_2,~V_{\text{e}}$ in Eqs.~(\ref{vdW01}) and~(\ref{Vfromb}) have relative errors up to $2\varsigma_1$ and $2\varsigma_2$, respectively, leading to relative errors of $2\varsigma_1$ for $\phi$ and $2\varsigma_2$ for $\pi-\alpha$ and $\theta$ in Eq.~(\ref{anglesI}) of Protocol I. Furthermore, incorrect laser Rabi frequency and timing also impact the accuracy of the Rabi pulse area, resulting in population leakage to Rydberg states unless their added effects cancel. During pulses-1 and 2 of Protocol I, the population transfer errors are about $(\pi \delta\Omega_k/2\Omega_k )^2$ and $(\pi \delta t/2t )^2$ for wrong Rabi frequencies and timing in a pulse duration of $t$, respectively. Because $(\pi \delta\Omega_k/2\Omega_k )^2\sim \varsigma_2^2$, these latter errors are negligible compared with those of the angles $\pi-\alpha$ and $\theta$. As a consequence, one needs the fluctuations of the laser phase and electric field bounded by $|\delta\beta_0/\beta_0|+|\delta\Omega_k/\Omega_k|\lessapprox10^{-3}$ to achieve the gate fidelity predicted in Fig.~\ref{figError}. The phase fluctuation of laser beams can be made much smaller than $10^{-3}$~\cite{Wineland1998,Saffman2005}, but the intensity fluctuation of lasers were several percent in typical experiments on Rydberg quantum gates~\cite{Gaetan2009,Wilk2010}. Nevertheless, lasers with root-mean-square intensity noise of less than $0.1\%$ were recently realized in preparing Rydberg states of $^{39}K$~\cite{Helmrich2017}, indicating a possibility to realize high-fidelity Barenco gates in near future. Alternatively, numerical simulation in~\cite{Goerz2014} showed that optimal control may be used to identify pulse sequences that are inherently robust to fluctuations of Rabi frequencies. Nevertheless, it is an open problem to implement these techniques in our Barenco gates to realize the gate performance of Fig.~\ref{figError}.

Severe atom loss during the gate sequence is another issue in current Rydberg gate experiments~\cite{Wilk2010,Isenhower2010,Zhang2010,Maller2015,Jau2015,Zeng2017}, and such loss may be from unwanted couplings that result in populating Rydberg states other than the targeted ones~\cite{Maller2015}. One possibility leading to unwanted couplings is level mixing due to stray electric fields~\cite{Saffman2016,Weiss2017}, which can be suppressed by microwave-induced dressing of Rydberg states~\cite{Booth2017}. Another possibility can be from multiple cycles of rise and fall of optical lasers~\cite{Maller2015}, a problem that may be partly avoided through exploiting gates that use only one pulse for qubit entanglement~\cite{PhysRevLett.85.2208,Han2016,Su2016,Su2017}.

\section{Conclusions}
In conclusion, we propose two protocols to realize a universal quantum gate $\mathbb{B}$ based on a tunable non-collinear interaction of the form $V|r_1r_2\rangle\langle r_1r_3|+$H.c.. We show that this non-collinear interaction is achievable for a quantum system that exhibits a usual blockade interaction of the form $\sum_{j, k}b_{jk}|R_jR_k\rangle\langle R_jR_k|$, such as Coulomb blockade in quantum dots or Rydberg blockade in neutral atoms. Among the three angles $\alpha,\theta$, and $\phi$ in $\mathbb{B}$, $\phi$ is freely tunable via adjusting phases of external fields in the first protocol, while the other two angles in the first protocol and all the three angles in the second protocol can be tuned by adjusting the interaction coefficients and wait durations. In particular, the first protocol can also lead to the CNOT and Controlled-Y gates. Analyses of the gate protocols by using Rydberg interaction in neutral atoms show that the gate operation time can be in the microsecond regime with intrinsic fidelity error on the order of $10^{-3}$. Such an intrinsic fidelity, however, is achievable only if technical problems do not occur, which is an open problem at the moment.

\section*{ACKNOWLEDGMENTS}
The author thanks Yan Lu for fruitful discussions and acknowledges support from the Fundamental Research Funds for the Central Universities and the 111 Project (B17035).

\appendix{}
  \section{Interactions in the example of $\mathbb{B}$ with neutral atoms}\label{appA}
We consider two-photon transitions from the ground states to Rydberg states via the intermediate hyperfine level $|5p_{1/2},F=1,m_F=1\rangle$, and choose $(n_1,n_2)=(96,102)$. Then, the vdWI coefficients are~\cite{Shi2014} $C_6(|R_0R_1\rangle)  =35.71\times2\pi$~THz~$\mu m^6$ and $C_6(|R_0R_2\rangle)  =-10.07\times2\pi$~THz~$\mu m^6$. There is an exchange interaction between $|R_0R_2\rangle$ and $|R_2R_0\rangle$ with a tiny vdWI coefficient $C_6' =-5\times2\pi$~GHz~$\mu m^6$, which can be neglected. The character of the interaction transfers between resonant dipole-dipole interaction and vdWI at about $5\mu$m. We choose a two-atom distance of $l=20\mu $m to guarantee the picture of vdWI. Then, $\{b_{01} , b_{02}\} = \{ 0.558,-0.157\}\times2\pi\text{MHz}$, and we have the following interaction strengths in units of $2\pi\text{kHz}$,
\begin{eqnarray}
  V_1&=& 558\cos^2\beta_1 -157 \sin^2\beta_1,\nonumber\\
  V_2&=& 558\sin^2\beta_1 -157 \cos^2\beta_1,\nonumber\\
  V_{\text{e}}&=& 715 \sin\beta_1 \cos\beta_1.\nonumber
\end{eqnarray} 
  
\section{Gate fidelity error}\label{appB}
The excited states of atoms can experience decay, the Rabi frequency $\Omega$ is not infinitely large compared with the blockade $V_{1(2)}$, and there can be population leakage out of the computational basis. The force between the two atoms can also induce drift of the atomic spacing. When the atoms are captured by optical dipole traps before and after the gate sequence, the distance between the two atoms can vary from the ideal $l$. These several factors cause error to the gate operation. We denote an input state by a wave-function $|\Psi_{\text{in}}\rangle$, and the output state by a density matrix $\rho_{\text{out}}$, which can be different from $\mathbb{B} |\Psi_{\text{in}}\rangle\langle \Psi_{\text{in}}|\mathbb{B}^\dag  $. Then the fidelity of the gate can be defined as
\begin{eqnarray}
\mathcal{F} &=& \overline{\langle \Psi_{\text{in}}|\mathbb{B}^\dag \rho_{\text{out}} \mathbb{B} |\Psi_{\text{in}}\rangle    },\nonumber
\end{eqnarray}
where the over-line means an average over all possible input states. Take Protocol I as an example, the Rydberg state decay, finiteness of $\Omega$, and population leakage lead to errors that can be respectively approximated by~\cite{Saffman2005,Shi2017pra}
\begin{eqnarray}
  E_{\text{de}}
  &=& \frac{( \frac{\pi}{\Omega}+T)(\tau_1+\tau_2)/2 }{ \tau_1\tau_2  + \sin^2\beta_1\cos^2\beta_1(\tau_1 -\tau_2)^2 }+( \frac{\pi}{2\Omega}+T/2)/\tau_1 ,\nonumber\\
  E_{\text{bl}} &=& 2\frac{V_1^2+V_2^2}{\Omega^2} ,\nonumber\\
 E_{\text{le}} &=&\frac{\Omega^2}{\Delta_1^2} + \frac{\Omega^2}{2\Delta_2^2},\nonumber
\end{eqnarray}
where $\tau_j$ is the lifetime of the state $|R_j\rangle$, $j=1,2$. Here, $\Delta_{1(2)}/2\pi=1.8(1.5)$~GHz is the detuning for the dominant leakage channel. As an example, the nearest levels to $|R_0\rangle$ that can be transferred from the $5p_{1/2}$ state is $| 94d_{3/2},m_J=1/2,m_I=3/2\rangle$, with a detuning of $\Delta_1/2\pi = 1.8$~GHz.

During the wait periods, because both atoms are in Rydberg states, entanglement between the motional states and the internal states may arise. This effect, however, is negligible. For example, if two atoms are in the state $|R_0R_1\rangle=|R_1R_1\rangle$, a force $-6C_6(|R_0R_1\rangle) /l^7$ arises. With this force, the relative speed between the two atoms changes by $|\delta v|=6C_6(|R_0R_1\rangle)T_{\text{Ry}} /\mu l^7$, where $\mu$ is the mass of the atom and $T_{\text{Ry}}$ is the time for the atoms to be in the Rydberg states. For $T_{\text{Ry}} =1\mu$s and $l=20\mu $m, we have $\delta v= 7.6\times10^{-4}m/s$. If the initial relative speed is zero, the two-atom separation will change by about $3.8\times10^{-4}\mu$m after one gate cycle, which is negligible compared with $l$.

For gate fidelity errors caused by distance fluctuation of the atoms, we note that the parameters characterizing a trap include the trap depth $U$, the oscillation frequencies $\{\omega_x,\omega_y,\omega_z\}$, and the averaged variances of the position $\{\sigma_x^2,\sigma_y^2,\sigma_z^2\}$.
We consider the case when the motional state of a trapped neutral atom is thermal, i.e., $k_BT_a/2\geq \hbar \omega_j$, $j=x,y,z$, where $k_B$ is the Boltzmann constant and $T_a$ is the effective temperature of atoms. For an optical tweezer created by a single laser beam with wavelength $\lambda$ and waist $w$ that propagates along $z$, we have $\sigma_x^2= \sigma_y^2 =\frac{w^2}{4}\frac{T_{a}}{U}$, $\sigma_z^2=\xi^2 \sigma_x^2$, $\xi = \sqrt2\pi w/\lambda$, where $U$ and $\xi$ are the potential depth and anisotropy factor of the trap, respectively~\cite{Saffman2005}. The position distributions of the two qubits depend on $\{\sigma_x^2,\sigma_y^2,\sigma_z^2\}$. In different runs of the gate, the fluctuation of the atomic location adds an extra error $\overline{E}_{\mathcal{L}}$ to the total gate fidelity error, which can be numerically evaluated by Monte Carlo integration~\cite{Shi2017pra}. For $\{w,\lambda \}=\{3.0,1.1\}\mu$m, $T=0.5\mu$s, and $U=20$~mK, numerical calculation shows that it is in the interval of $[1.4,~52]([1.0,~49])\times10^{-4}$ for Protocol I~(II) when $T_a\in[10,200]\mu$K. We also considered a similar set-up analyzed in Ref.~\cite{Shi2017} if atoms are cooled to motional ground states, and analyzed $\overline{E}_{\mathcal{L}}$ as a function of trap depth $U$. Numerical calculation shows that $\overline{E}_{\mathcal{L}}<7\times10^{-5}$ for both protocols when $U>1\mu$K. These analyses mean that one can suppress the error caused by position fluctuation through laser cooling of atoms.

%

\end{document}